\documentclass{article}
\usepackage{witmse}
\usepackage{graphics}   
\usepackage{amsmath}
\usepackage{amssymb}
\usepackage{amsthm}
\newtheorem{theorem}{Theorem}

\newtheorem{proposition} {Proposition}

\newtheorem{algorithm}{Algorithm}

\title{QUANTIZATION OF DISCRETE PROBABILITY DISTRIBUTIONS\footnotemark}

\name{Yuriy A. Reznik}
\address{Qualcomm Inc.   \\
  5775 Morehouse Drive,
  San Diego, CA 92121 \\
  yreznik@ieee.org }
\begin{document}

\maketitle
\addtocounter{footnote}{1}
\footnotetext{Presented at the Information Theory and Applications (ITA) workshop, San Diego, CA, February 1--5, 2010, and at the Workshop on Information Theoretic Methods in Science and Engineering (WITMSE), Tampere, Finland, August 16 -- 18, 2010.}

\begin{abstract}
We study the problem of quantization of discrete probability distributions, arising in universal coding, as well as other applications. We show, that in many situations this problem can be reduced to the covering problem for the unit simplex. Such setting yields precise 
asymptotic characterization in the high-rate regime. Our main contribution is a simple and asymptotically optimal algorithm for solving this problem. Performance of this algorithm is studied and compared with several known solutions.
\end{abstract}

\section{INTRODUCTION}
\label{sec:intro}
The problem of coding of probability distributions surfaces many times in the history of source coding. First universal codes, developed in late $1960$s, such as {\em Lynch-Davis\-son\/}~\cite{Lynch66,Davisson66}, {\em combinatorial\/}~\cite{ShtarkovBabkin71}, and {\em enumerative\/} codes~\cite{Cover73} used lossless encoding of frequencies of symbols in an input sequence. The {\em Rice machine\/}~\cite{RicePlaunt}, developed in early $1970$'s, transmitted quantized estimate of variance of source's distribution. {\em Two-step universal codes\/}, developed by J.~Rissanen in $1980$s, explicitly estimate, quantize, and transmit parameters of distribution as a first step of the encoding process \mbox{\cite{Rissanen84,Rissanen96}}. Vector quantization techniques for two-step universal coding were proposed in~\cite{ZegerBistLinder,ChouEffrosGray}.
In practice, two-step coding was often implemented by constructing a Huffman tree, then encoding and transmitting this code tree, and then encoding and transmitting the data. Such algorithms become very popular in $1980$s and $1990$s, and were used, for example, in ZIP archiver~\cite{PKZIP}, and JPEG image compression standard~\cite{JPEG}.

In recent years, the problem of coding of distributions has attracted a new wave of interest coming from other fields. For example, in computer vision, it is now customary to use histogram-derived descriptions of image features. Examples of such descriptors include SIFT~\cite{sift}, SURF~\cite{surf}, and CHoG~\cite{CHoG}, differentiating mainly in a way the quantize histograms. Several other uses of coding of distributions are described in~\cite{Gagie}.


To the best of our knowledge, most related prior studies were motivated by optimal design of universal source codes \cite{Rissanen84}, \cite{Rissanen96}, \cite{ZegerBistLinder}, \cite{ChouEffrosGray}, \cite{HanKobayashi}. In this context, quantization of distributions becomes a small sub-problem in a complex rate optimization process, and final solutions yield very few insights about it.

In this paper, we study quantization of distributions as a stand-alone problem.
%
%
%
%
%
In Section~2, we introduce notation and formulate the problem. In Section~3, we study achievable performance limits. In Section~4, we propose and study an algorithm for solving this problem. In Section~5, we provide comparisons with other known techniques. Conclusions are drawn in Section~6.

\section{DESCRIPTION OF THE PROBLEM}
Let $A=\{r_1,\ldots,r_m\}$, $m < \infty$, denote a discrete set of events, and let $\Omega_m$ denote the set of probability distributions over $A$:
\begin{equation}
\Omega_m = \left\{{ [\omega_1,\ldots,\omega_m] \in \mathbb{R}^m \Bigl\vert\Bigr.\, 
\omega_i \geqslant 0\,, \textstyle \sum_i \omega_i = 1}\right\}.
\end{equation}
Let $p \in \Omega_m$ be an input distribution that we need to encode, and let $Q \subset \Omega_m$ be a set of distributions that we will be able to reproduce. We will call elements of $Q$ {\em reconstruction points\/} or {\em centers} in $\Omega_m$. We will further assume that $Q$ is finite $|Q| < \infty$, and that its elements are enumerated and encoded by using fixed-rate code. The rate of such code is $R(Q) = \log_2 |Q|$ bits.
By~$d\,(p,q)$ we will denote a {\em distance measure\/} between distributions $p,q \in \Omega_m$.

In order to complete traditional setting of the quantization problem for distribution $p \in \Omega_m$, it remains to assume that it is produced by some random process, e.g. a memoryless process with density $\theta$ over $\Omega_m$. Then the problem of quantization can be formulated as minimization of the {\em average distance} to the nearest reconstruction point~(cf.~\cite[Lemma~3.1]{GrafLuschgy})
    \begin{equation}
    \bar{d}(\Omega_m,\theta,R) = \inf_{\substack{ Q \subset \Omega_m \\ |Q| \leqslant 2^R}} \mathbf{E}_{\substack{p \in \Omega_m \\ p \sim \theta~~}} ~\min_{q \in Q} d(p,q)\,, \label{ee:d_bar}
    \end{equation}


However, we notice that in most applications, best possible accuracy of the reconstructed distribution is needed {\em instantaneously\/}.  For example, in the design of a two-part universal code, sample-derived distribution is quantized and immediately used for
encoding of this sample~\cite{Rissanen96}. Similarly, in computer vision / image recognition applications, histograms of gradients from an image are extracted, quantized, and used right away to find nearest match for this image.

In all such cases, instead of minimizing the expected distance, it makes more sense to design a quantizer that minimizes the {\em worst case-\/} or {\em maximal distance\/} to the nearest reconstruction point. In~other words, we need to solve the following problem
\footnote{
The dual problem
$$
R(\varepsilon) = \inf_{{Q \subset \Omega_m: \max_{p\in\Omega_m} \min_{q \in Q} d(p,q) \leqslant \varepsilon}} \log_2 |Q|\,,
$$
may also be posed. The resulting quantity $R(\varepsilon)$ can be understood as  Kolmogorov's \mbox{\em $\varepsilon$-entropy\/} for metric space $(\Omega_m,d)$~\cite{KolmogorovTikhomirov}.}
    \begin{equation}
    d^{*}(\Omega_m,R) = \inf_{\substack{ Q \subset \Omega_m\\|Q| \leqslant 2^R}} ~\max_{p\in\Omega_m} ~\min_{q \in Q} d(p,q)\,. \label{ee:d_star}
    \end{equation}
We next survey some known results about it.

\section{ACHIEVABLE PERFORMANCE LIMITS}
We note that the problem (\ref{ee:d_star}) is purely geometric in nature.
It is equivalent to the {\em problem of covering of $\Omega_m$
with at most $2^R$ balls of equal radius\/}. Related and immediately applicable results can be found in Graf and Luschgy~\cite[Chapter 10]{GrafLuschgy}.

First, observe that $\Omega_m$ is a compact set in $\mathbb{R}^{m-1}$ (unit $m-1$-simplex), and that its volume 
in $\mathbb{R}^{m-1}$ can be computed as follows~\cite{Sommerville}:
\begin{equation}
\lambda^{m-1}(\Omega_m) = \left.\frac{a^k}{k!}\sqrt{\frac{k+1}{2^k}}~\right\vert_{\substack{k=m-1 \\ a=\sqrt{2}}} = \frac{\sqrt{m}}{(m-1)!}\,.
\end{equation}
Here and below we assume that $m \geqslant 3$.


Next, we bring result for asymptotic covering radius \cite[Theorem~10.7]{GrafLuschgy}
\begin{equation}
\lim_{R \rightarrow \infty} 2^{\frac{R}{m-1}} d^{*}(\Omega_m,R) = C_{m-1} \!\! \sqrt[m-1]{\lambda^{m-1} (\Omega_m)},
\end{equation}
where $C_{m-1}>0$ is a constant known as {\em covering coefficient for the unit cube\/}
\begin{equation}
C_{m-1} = \inf_{R \geqslant 0} 2^{\frac{R}{m-1}} d^{*}([0,1]^{m-1},R).
\end{equation}

The exact value of $C_{m-1}$ depends on a distance measure $d(p,q)$. For example, for~$L_\infty$ norm
\begin{equation*}
d_{\infty}(p,q) =||p-q||_{\infty} = \max_i |p_i - q_i|\,,
\end{equation*}
it is known that 
\begin{equation*}
C_{m-1,\infty} = \tfrac{1}{2}\,.
\end{equation*}
Hereafter, when we work with specific $L_{r}$ - norms:
\begin{equation}
d_r(p,q) = ||p-q||_{r} = \biggl(\sum_i |p_i - q_i|^{r} \biggr)^{1/r} 
\end{equation}
we will attach subscripts $r$ to covering radius $d^*(.)$ and other expressions to indicate type of norm being used.
%

By putting all these facts together, we obtain:
\begin{proposition}
With $R \rightarrow \infty$:
\begin{equation}
 d_{\infty}^{*}(\Omega_m,R) \sim  \tfrac{1}{2} \sqrt[m-1]{\tfrac{\sqrt{m}}{(m-1)!}} ~2^{-\frac{R}{m-1}} \label{eq:d_star_infty}
\end{equation}
and more generally (for other $L_r$-norms, $r \geqslant 1$):
\begin{equation}
 d_{r}^{*}(\Omega_m,R) \sim  C_{m-1,r} \sqrt[m-1]{\tfrac{\sqrt{m}}{(m-1)!}} ~2^{-\frac{R}{m-1}}, \label{eq:d_star_alpha}
\end{equation}
where $C_{m-1,r}$ are some constants.
\end{proposition}


We further note, that with large $m$ the leading term in~(\ref{eq:d_star_alpha}) turns into
\begin{equation}
\sqrt[m-1]{\tfrac{\sqrt{m}}{(m-1)!}} = \frac{e}{m} + O\left(\frac{1}{m^2}\right)
\end{equation}
which is a decaying function of $m$. This highlights an interesting property and distinction of the problem of quantization of $m$-ary distributions, compared to quantization of the unit $(m-1)$-dimensional cube.


Our next task is to design a practical algorithm for solving this problem.

\section{PRACTICAL ALGORITHM FOR CODING OF DISTRIBUTIONS}

\subsection{Algorithm design}

Our algorithm can be viewed as a custom designed lattice quantizer.
It is interesting in a sense that its lattice coincides with the concept of {\em types\/} in universal coding~\cite{Csiszar1998}.

\subsubsection{Type Lattice}
Given some integer $n\geqslant 1$, define a lattice:
\begin{equation}
Q_n = \left\{ [q_1,\ldots,q_m] \in \mathbb{Q}^m \Bigl\vert\Bigr. \textstyle q_i = \frac{k_i}{n}, \sum_i k_i = n  \right\},\!\! \label{eq:type}
\end{equation}
where $n, k_1,\ldots,k_m \in \mathbb{Z}^{+}$. Parameter $n$ serves as a common denominator to all fractions, and can be used to control the density and number of points in $Q_n$.

\begin{figure}
\centering
\resizebox{1.in}{!}{\includegraphics{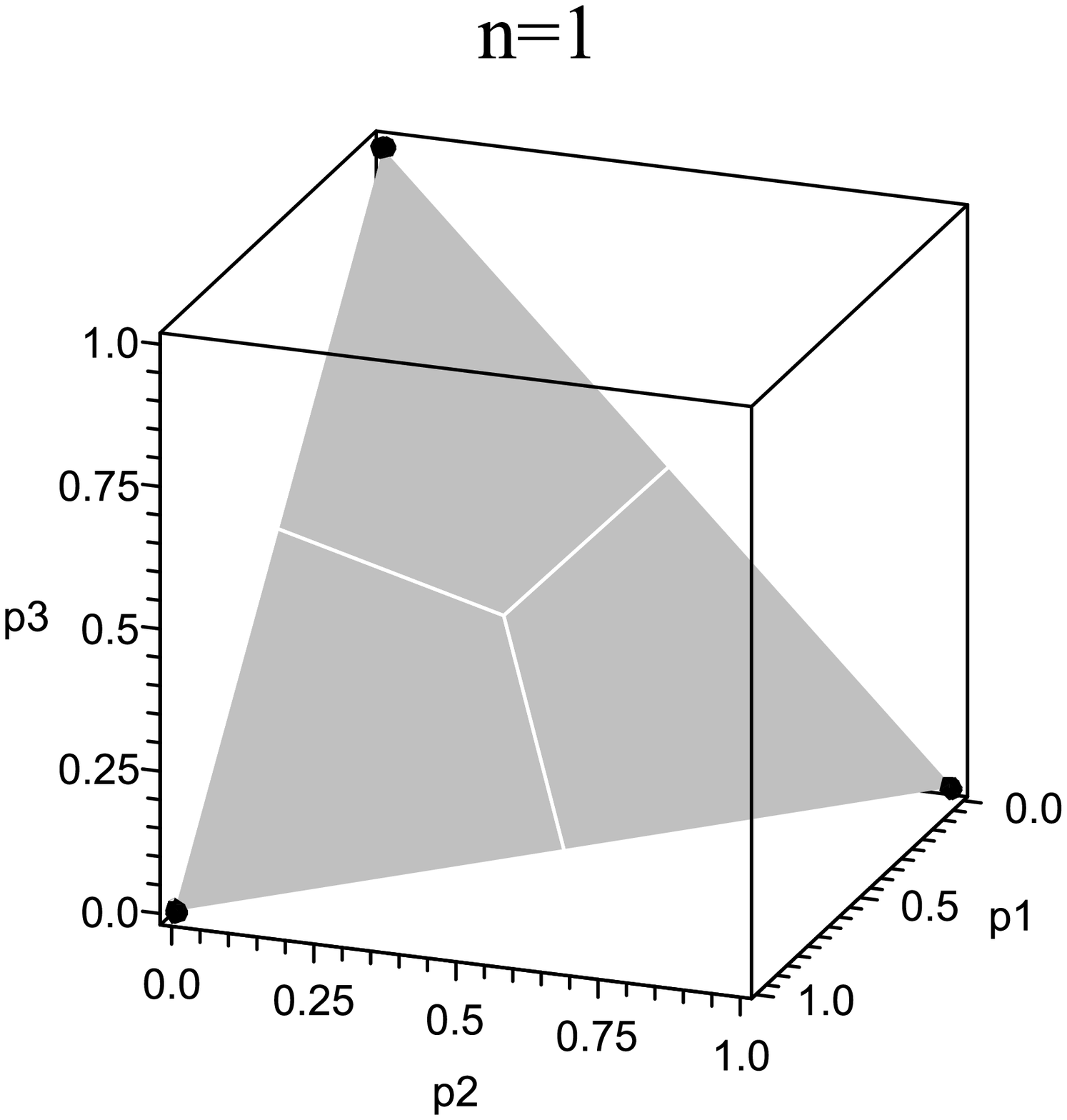}}\,
\resizebox{1.in}{!}{\includegraphics{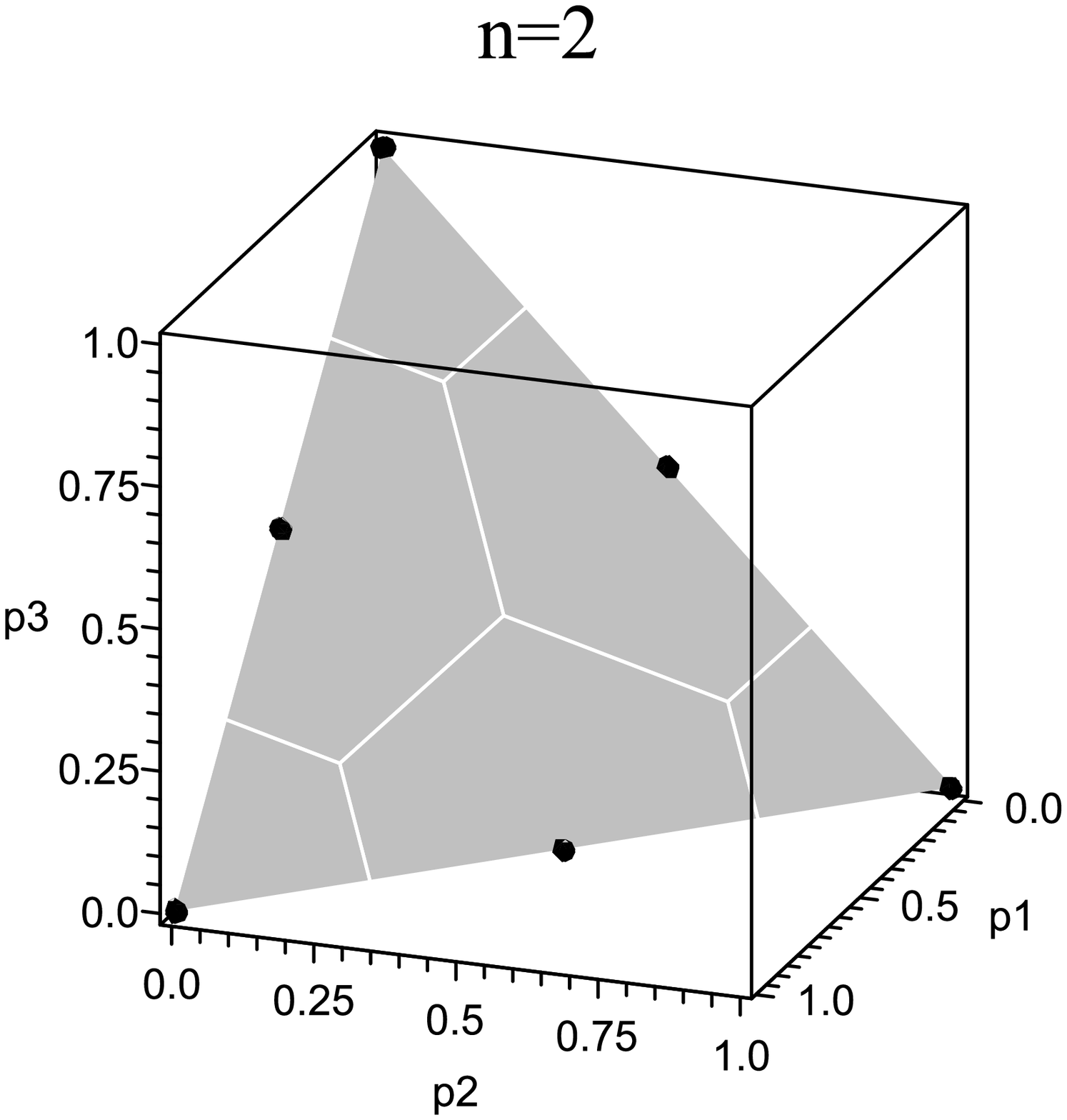}}\,
\resizebox{1.in}{!}{\includegraphics{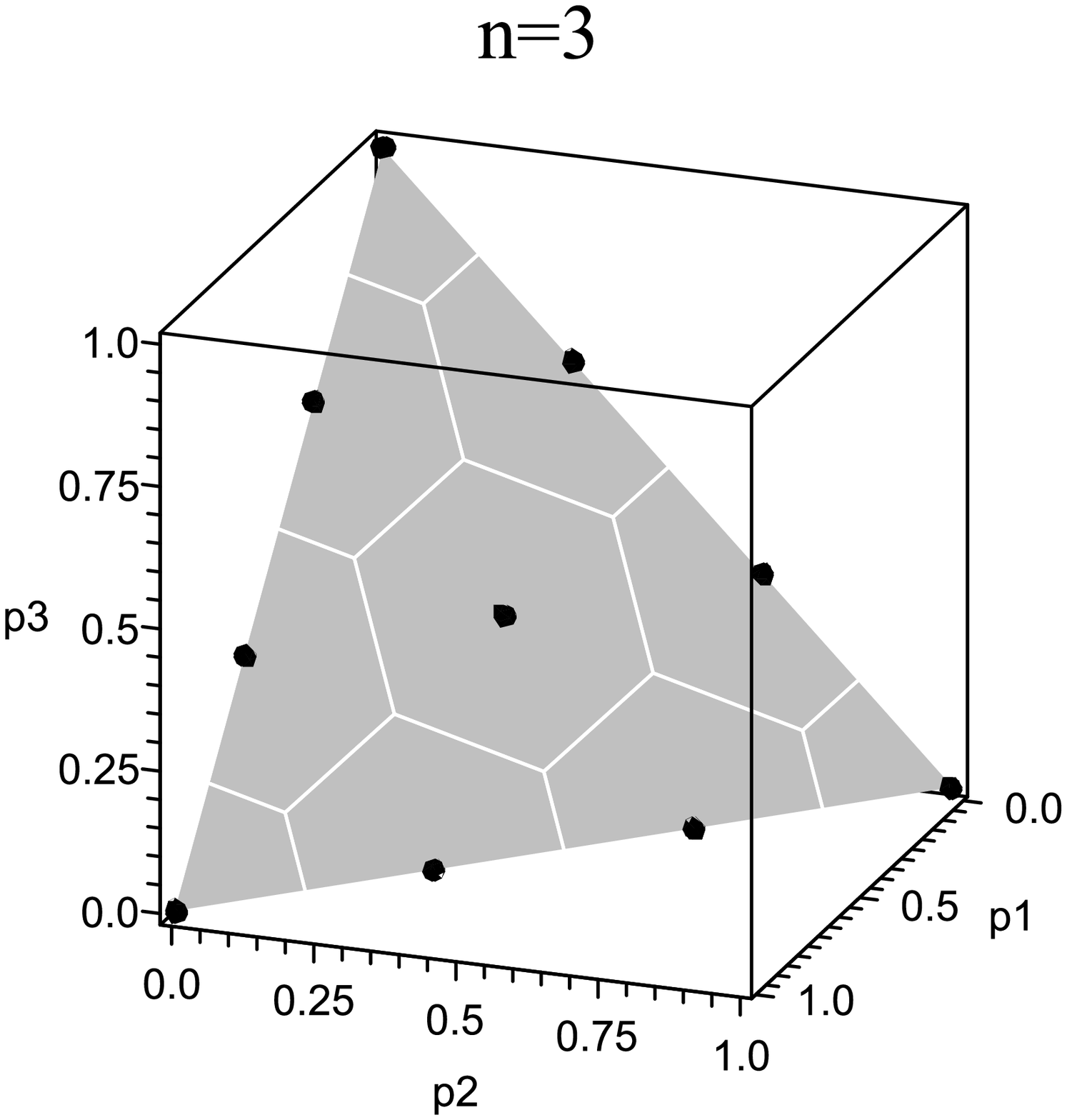}}
\caption{Examples of type lattices and their Voronoi cells in $3$ dimensions $(m=3,n=1,2,3)$.}
\end{figure}

By analogy with the concept of {\em types\/} in universal coding~\cite{Csiszar1998} we will refer to distributions $q \in Q_n$ as types. For~same~reason we will call $Q_n$ a {\em type lattice\/}.
Several examples of type lattices are shown in Figure~1.

\subsubsection{Quantization}
The task of finding the nearest type in $Q_n$ can be solved by using the following simple algorithm:
\footnote{This algorithm is similar in concept to Conway and Sloane's quantizer for lattice $A_n$~\cite[Chapter~20]{SPLAG}. It works within $(m$-$1)$ simplex.}:
\begin{algorithm}
Given $p,n$, find nearest $q = \left[\tfrac{k_1}{n},\ldots,\tfrac{k_m}{n}\right]$:
\begin{enumerate}
\setlength{\topsep}{0in} \setlength{\parskip}{0in} \setlength{\parsep}{0in} \setlength{\itemsep}{0in}
\item Compute values ($i=1,\ldots,m$)
\begin{equation*}
k'_i = \left \lfloor n p_i + \tfrac{1}{2}\right \rfloor\,, ~~ \textstyle n' = \sum_i k'_i\,.
\end{equation*}
\item If $n' = n$ the nearest type is given by:
$k_i = k'_i$.
Otherwise, compute errors
\begin{equation*}
\delta_i =  k'_i - n p_i\,,
\end{equation*}
and sort them such that
\begin{equation*}
-\tfrac{1}{2} \leqslant \delta_{j_1} \leqslant \delta_{j_2} \leqslant \ldots  \leqslant \delta_{j_m} \leqslant  \tfrac{1}{2} \,,
\end{equation*}
\item Let $\Delta = n' - n$. If $\Delta>0$ then decrement $d$ values $k'_i$ with largest errors
\begin{equation*}
k_{j_i} = \left[
\begin{array}{ll}
k'_{j_i} & j = i, \ldots, m - \Delta -1\,,\\
k'_{j_i} - 1 & i = m-\Delta,\ldots, m\,,
\end{array}
\right.
\end{equation*}
otherwise, if $\Delta < 0$ increment $|\Delta|$ values $k'_i$ with smallest errors
\begin{equation*}
k_{j_i} = \left[
\begin{array}{ll}
k'_{j_i} + 1 & i = 1, \ldots, |\Delta|\,,\\
k'_{j_i} & i = |\Delta|+1,\ldots, m\,.
\end{array}
\right.
\end{equation*}
\end{enumerate}
\end{algorithm}
The logic of this algorithm is obvious. It~finds points that are nearest in terms of L-norms.
By~using quick-select instead of full sorting in step~2, its run time can be reduced to $O(m)$.


\subsubsection{Enumeration and Encoding}
As mentioned earlier, the number of types in lattice $Q_n$ depends on the parameter $n$.
It is essentially the number of partitions of $n$ into $m$ terms $k_1+\ldots+k_m =n$:
\begin{equation}
|Q_n| = \binom{n+m-1}{m-1}\,. \label{eq:xi}
\end{equation}

In order to encode a type with parameters $k_1,\ldots,k_m$, we need to obtain its unique index $\xi(k_1,\ldots,k_m)$. 
We~suggest to compute it as follows:
\begin{eqnarray}
\lefteqn{\xi(k_1,\ldots,k_n) = }  \label{eq:f_map} \\
& & \sum^{n-2}_{j=1} \sum_{i=0}^{k_j-1} \binom{n- i - \sum_{\,l=1}^{j-1} k_l + m-j-1}{m-j-1} + k_{n-1}. \nonumber
\end{eqnarray}
This formula follows by induction (starting with $m=2,3$, etc.), and it implements lexicographic enumeration of types. For example:
\begin{eqnarray*}
    \xi(0,0,\ldots,0,n)   & = & 0\,, \\
    \xi(0,0,\ldots,1,n-1) & = & 1\,, \\
    \ldots & & \\
    \xi(n,0,\ldots,0,0)   & = & \textstyle \binom{n+m-1}{m-1}-1\,.
\end{eqnarray*}
Similar combinatorial enumeration techniques were discussed in~\cite{ShtarkovBabkin71,Cover73,Schalkwijk72}. With precomputed array of binomial coefficients, the computation of an index by using this formula requires only $O(n)$ additions\footnote{Considering that $\log |Q_n| = O(\log n)$, this translates into at most $O(n \log n)$ bit-additions}.

Once index is computed, it is transmitted by using its direct binary representation at rate:
\begin{equation}
R(n) = \left\lceil \log_2 \textstyle \binom{n+m-1}{m-1} \right\rceil\,. \label{eq:type_rate}
\end{equation}

\subsection{Analysis}
Type lattice $Q_n$ is related to so-called {\em lattice $A_n$\/} in lattice theory~\cite[Chapter~4]{SPLAG}.
It can be understood as a bounded subset of $A_n$ with $n=m-1$ dimensions, which is subsequently scaled, and placed in the unit simplex.

Using this analogy, we can show that vertices of Voronoi cells (so called {\em holes\/}) in type lattice $Q_n$ are located at:
\begin{equation}
q_i^* = q + v_i, ~~~q \in Q_n, ~~~ i=1,\ldots,m-1,
\end{equation}
where $v_i$ are so-called {\em glue vectors\/}~\cite[Chapter~21]{SPLAG}:
\begin{equation}
v_i = \tfrac{1}{n} \biggl[\underbrace{\tfrac{m-i}{m},\ldots,\tfrac{m-i}{m}}_{i~\mathrm{times}},~ \underbrace{\tfrac{-i}{m},\ldots,\tfrac{-i}{m}}_{m-i~\mathrm{times}}\biggr]\,. \label{eq:glue_vectors}
\end{equation}

We next compute maximum distances (covering radii).
\begin{proposition}
Let $a = \left\lfloor m/2\right\rfloor$.
The following holds:
\begin{eqnarray}
\max_{p\in\Omega_m} \min_{q \in Q_n} d_{\infty}(p,q) & = & \tfrac{1}{n} \left(1 - \tfrac{1}{m}\right)\,, \label{eq:r_inf}\\
\max_{p\in\Omega_m} \min_{q \in Q_n} d_{2}(p,q) & = & \tfrac{1}{n} \sqrt{\tfrac{a(m-a)}{m}}\,, \label{eq:r_2}\\
\max_{p\in\Omega_m} \min_{q \in Q_n} d_{1}(p,q) & = & \tfrac{1}{n} \tfrac{2a(m-a)}{m}\,. \label{eq:r_1}
\end{eqnarray}
\end{proposition}
\begin{proof}
We use vectors (\ref{eq:glue_vectors}). The largest component values appear when $i=1$ or $i=m-1$.
E.g. for $i=1$:
\begin{equation*}
v_1 = \tfrac{1}{n} \left[\tfrac{m-1}{m},\tfrac{-1}{m},\ldots,\tfrac{-1}{m}\right]\,.
\end{equation*}
This produces $L_\infty$ - radius.
The largest absolute sum is achieved when all components are approximately the same in magnitude.
This happens when $i=a$:
\begin{equation*}
v_a = \tfrac{1}{n} \biggl[\underbrace{\tfrac{m-a}{m},\ldots,\tfrac{m-a}{m}}_{a~\mathrm{times}}, ~ \underbrace{\tfrac{-a}{m},\ldots,\tfrac{-a}{m}}_{m-a~\mathrm{times}}\biggr]\,.
\end{equation*}
This produces $L_1$ -  radius.
$L_2$ norm is the same for all vectors $v_i, i>0$.
\end{proof}

It remains to evaluate distance / rate characteristics of type-lattice quantizer:
\begin{equation*}
d^*_{r}[Q_n](\Omega_m,R) =  \displaystyle \min_{n: |Q_n| \leqslant 2^R} \max_{p\in\Omega_m} \min_{q \in Q_n} d_{r}(p,q) \,.
\end{equation*}
We report the following.
\begin{theorem}
Let $a = \left\lfloor m/2\right\rfloor$.
Then, with $R \rightarrow \infty$:
\begin{eqnarray}
d^*_{\infty}[Q_n](\Omega_m,R) & \sim & \displaystyle 2^{-\frac{R}{m-1}} \frac{1 - \tfrac{1}{m}}{\sqrt[m-1]{\left(m-1\right)!}}  \,, \\
d^*_{2}[Q_n](\Omega_m,R) & \sim & \displaystyle 2^{-\frac{R}{m-1}} \frac{\sqrt{\tfrac{a(m-a)}{m}}}{\sqrt[m-1]{\left(m-1\right)!}}  \,, \\
d^*_{1}[Q_n](\Omega_m,R) & \sim & \displaystyle 2^{-\frac{R}{m-1}} \frac{\tfrac{2a(m-a)}{m}}{\sqrt[m-1]{\left(m-1\right)!}}\,.
\end{eqnarray}
\end{theorem}
\begin{proof}
We first obtain asymptotic (with $n \rightarrow \infty$) expansion for the rate of our code (\ref{eq:type_rate}):
\begin{equation*}
R = (m-1) \log_2 n - \log_2 \left(m-1\right)! + O\left(\tfrac{1}{n}\right)\,.
\end{equation*}
This implies that
\begin{equation*}
n \sim 2^{\frac{R}{m-1}} \sqrt[{m-1}]{(m-1)!}\,.
\end{equation*}
Statements of theorem are obtained by combination of this relation with expressions (\ref{eq:r_inf}-\ref{eq:r_1}).
\end{proof}

\subsubsection{Optimality}
We now compare the result of Theorem~1 with theoretical asymptotic estimates for covering radius for $\Omega_m$ (\ref{eq:d_star_infty}, \ref{eq:d_star_alpha}). As~evident, the maximum distance in our scheme decays with the rate~$R$~as:
\begin{equation*}
d^*[Q_n](\Omega_m,R) \sim 2^{-\frac{R}{m-1}},
\end{equation*}
which matches the decay rate of theoretical estimates.

The only difference is in a constant factor. For example, under $L_{\infty}$ norm, such factor in  expression (\ref{eq:d_star_infty}) is
\begin{equation*}
\frac{1}{2}  \sqrt[m-1]{\sqrt{m}} = \frac{1}{2} + O\left(\frac{\log m}{m} \right).
\end{equation*}
Our algorithm, on the other hand, uses a factor
\begin{equation*}
\frac{1}{2} \leqslant 1-\frac{1}{m} < 1,
\end{equation*}
which starts with $\frac{1}{2}$ when $m=2$.
This suggests that even in terms of leading constant our algorithm is close to the optimal.

\subsubsection{Performance in terms of KL-distance}
All previous results are obtained using L-norms. Such distance measures are common in computer vision applications~\cite{sift,gloh,surf}. In source coding, main interest presents Kullback-Leibler (KL) distance:
\begin{equation}
d_\mathrm{KL}(p,q) = D(p||q) = \sum_i p_i \log_2 \frac{p_i}{q_i} \,.
\end{equation}
It is not a true distance, so the exact analysis is complicated. Yet, by using Pinsker inequality~\cite{Pinsker1964}
\begin{equation}
d_\mathrm{KL}(p,q) \geqslant \tfrac{1}{2 \ln 2} \, d_1(p,q)^2\,, \label{eq:pinsker}
\end{equation}
we can at least show that for deep holes
\begin{equation*}
d_{\mathrm{KL}}(q^*,q) \geqslant \tfrac{1}{2\ln 2} \left( \tfrac{1}{n} \tfrac{2a(m-a)}{m} \right)^2\,.
\end{equation*}
By translating $n$ to bitrate, we obtain
\begin{equation}
d_{\mathrm{KL}}(q^*,q) \gtrsim \tfrac{1}{2\ln 2} ~ 2^{-\frac{2R}{m-1}} \biggl( \frac{\tfrac{2a(m-a)}{m}}{\sqrt[{m-1}]{(m-1)!}} \biggr)^2\,.
\end{equation}
More precise bounds can be obtained by using inequalities described in~\cite{fedorov}.

\subsection{Additional improvements}

\subsubsection{Introducing bias}
As easily observed, type lattice $Q_n$ places reconstruction points with $k_i = 0$ precisely on edges of the probability simplex $\Omega_m$. This is not best placement from quantization standpoint, particularly when $n$ is small. This placement can be improved by using {\em biased types\/}:
\begin{equation*}
q_i = \frac{k_i + \beta}{n + \beta m}\,,~~i=1,\ldots,m\,, \label{eq:type_with_prior}
\end{equation*}
where $\beta \geqslant 0$ is a constant that defines shift towards the middle of the simplex. In traditional source coding applications, it is customary to use $\beta=1/2$~\cite{KrichevskyTrofimov}.
In our case, setting $\beta=1/m$ appears to work best for $L$-norms, as it introduces same distance to edges of the simplex as covering radius of the lattice.

Algorithm~1 can be easily adjusted to find nearest points in such modified lattice.

\subsubsection{Using dual type lattice $Q_n^*$}
Another idea for improving performance of our quantization algorithm -- is to define and use {\em dual type lattice\/} $Q_n^*$. Such a lattice consists of all points:
\begin{equation*}
q^* = q + v_i, ~~q \in Q_n, ~~q^* \in \Omega_m ~~ i=0,\ldots,m-1,
\end{equation*}
where $v_i$ are the glue vectors (\ref{eq:glue_vectors}).

The main advantage of using dual lattice would be thinner covering at high dimensions (cf.~\cite[Chapter~2]{SPLAG}).
But even at small dimensions, it may sometimes be useful.
An~example of this for $m=3$ is shown in Figure.~2.

\begin{figure}
\label{fig0}
\centering
\resizebox{1.2in}{!}{\includegraphics{s-t3.eps}}
\resizebox{1.2in}{!}{\includegraphics{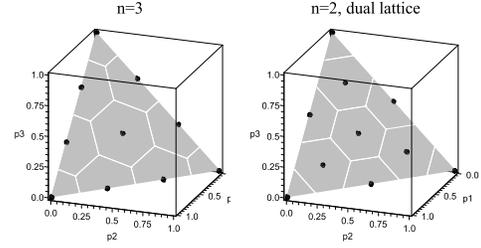}}
\caption{Two $10$-point lattices: $Q_3$ (left), and $Q^*_2$ (right). The second has much smaller cells.}
\end{figure}

\section{COMPARISON WITH OTHER TECHNIQUES}
Given a probability distribution $p \in \Omega_m$, one popular in practice way of compressing it is to design a prefix code (for example, a Huffman code) for this distribution $p$ first, and then encode the binary tree of such a code. Below we summarize some known results about performance of such schemes.

\subsection{Performance of tree-based quantizers}

\begin{figure*}
\centerline{\resizebox{2.7in}{!}{\includegraphics{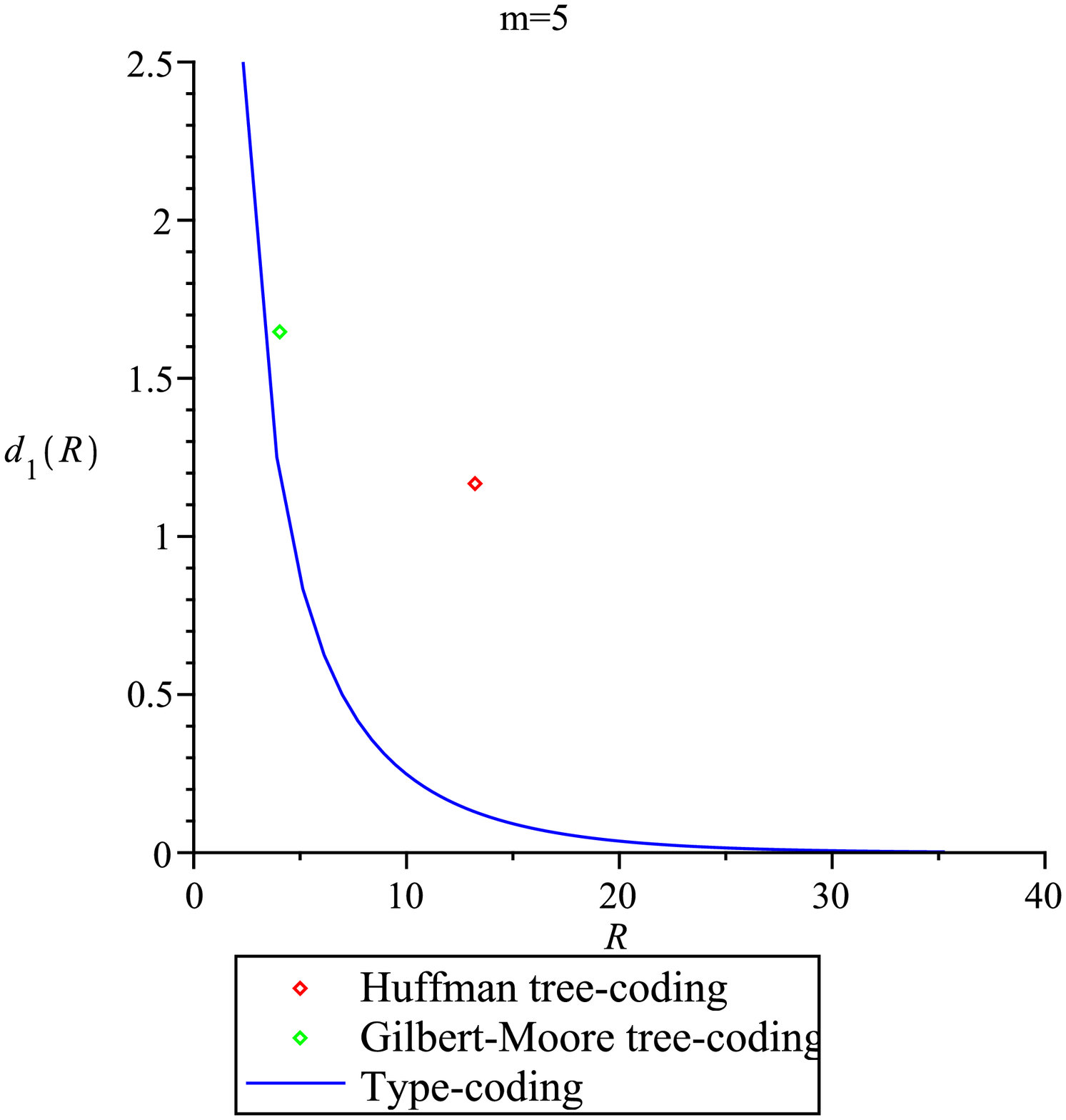}}~~~~~~~~~~~~~~
\resizebox{2.7in}{!}{\includegraphics{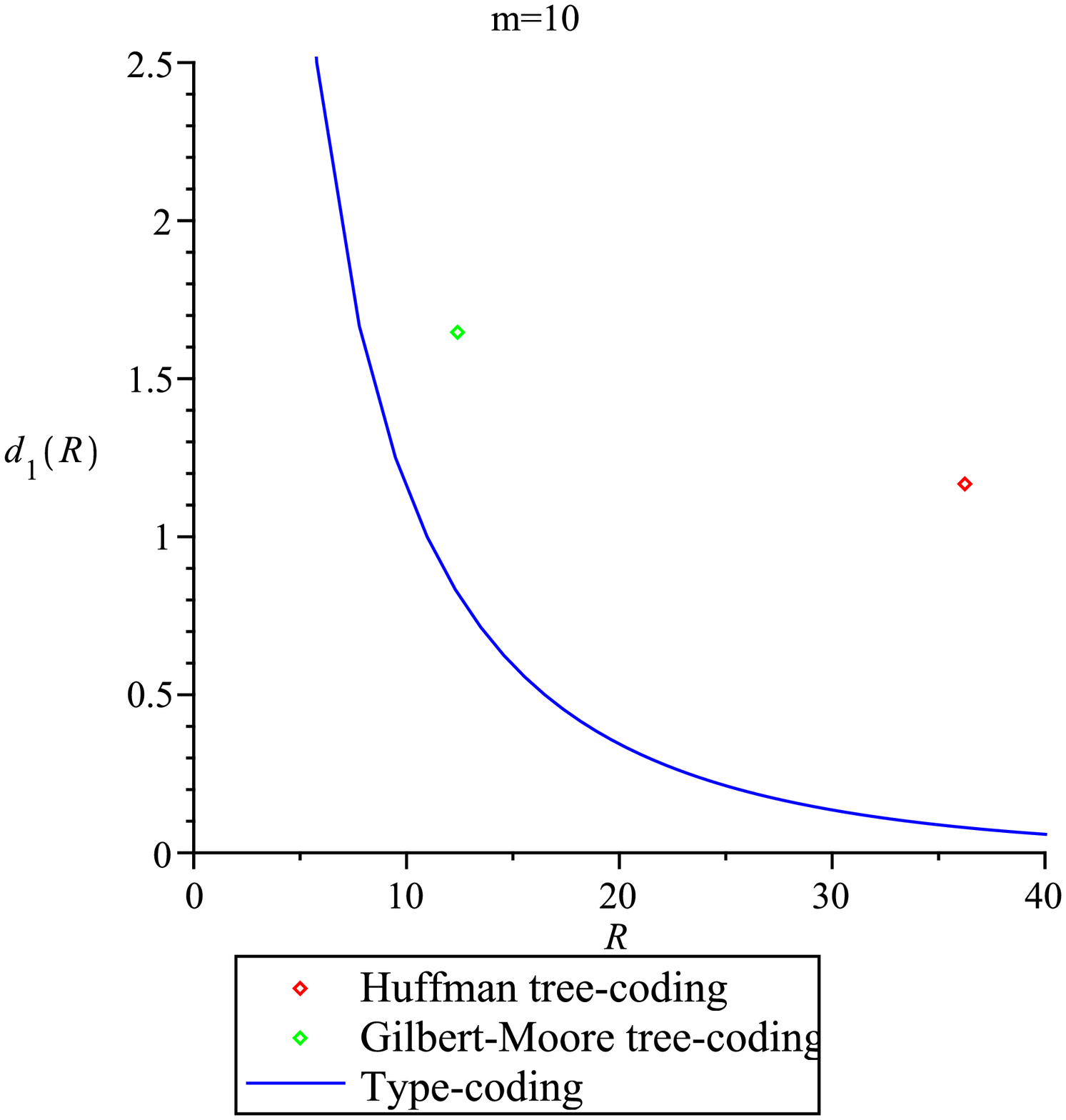}}~~~~~~}
\caption{Maximal $L_1$ distances vs rate characteristics $d^*_{1}[{\mathrm{H}}](R)$, $d^*_{1}[{\mathrm{GM}}](R)$, $d^*_{1}[{\mathrm{Q_n}}](R)$ achievable by Huffman-, Gilbert-Moore-, and type-based quantization schemes.} \label{fig:fig1}
\end{figure*}

By denoting by $\ell_1,\ldots,\ell_m$ lengths of prefix codes, recalling that they satisfy Kraft inequality~\cite{CoverThomas}, and noting that $2^{-\ell_i}$ can be used to map lengths back to probabilities, we arrive at the following set:
\begin{equation*}
Q_\mathrm{tree} = \left\{ [q_1,\ldots,q_m] \in \mathbb{Q}^m \bigl\vert\bigr. q_i = 2^{-\ell_i}, \textstyle \sum_i 2^{-\ell_i} \leqslant 1 \right\}.
\end{equation*}

There are several specific algorithms that one can employ for construction of codes, producing different subsets of $Q_\mathrm{tree}$. Below we only consider the use of classic Huffman and Gilbert-Moore~\cite{GilbertMoore} codes. Some additional tree-based quantization schemes can be found in~\cite{Gagie}.

\begin{proposition}
There exists a set $Q_{\mathrm{GM}} \subset Q_\mathrm{tree}$, such that
\begin{eqnarray}
d^{*}_{\mathrm{KL}}[Q_{\mathrm{GM}}](R_\mathrm{GM}) & \leqslant & 2\,, \label{eq:gm_red}\\
d^{*}_{1}[Q_{\mathrm{GM}}](R_\mathrm{GM}) & \leqslant & 2\sqrt{\ln 2}\,, \label{eq:gm_reg_l1} \\
d^{*}_{\infty}[Q_{\mathrm{GM}}](R_\mathrm{GM})  & \leqslant & 1\,,
\end{eqnarray}
where
\begin{eqnarray}
R_{\mathrm{GM}} = \log_2 |Q_{\mathrm{GM}}| & = & \log_2 C_{m-1} \\
& = & 2\,m - \tfrac{3}{2}\log_2 m + O(1), \nonumber
\end{eqnarray}
where $C_{n}=\frac{1}{n+1} \binom{2n}{n}$ is the Catalan number.
\end{proposition}
\begin{proof}
We use Gilbert-Moore code~\cite{GilbertMoore}. Upper bound for KL-distance is well known~\cite{GilbertMoore}. $L_{1}$ bound follows by Pinsker's inequality (\ref{eq:pinsker}).
$L_{\infty}$ bound is obvious: $p_i,q_i \in (0,1)$.
Gilbert-Moore code uses fixed assignment (e.g. from left to right) of letters to the codewords. Any binary rooted tree with $m$ leaves can serve as a code. The number of such trees is given by the Catalan number $C_{m-1}$.
\end{proof}

\begin{proposition}
There exists a set $Q_{\mathrm{H}} \subset Q_\mathrm{H}$, such that
\begin{eqnarray}
d^{*}_{\mathrm{KL}}[Q_{\mathrm{H}}](R_\mathrm{H}) & \leqslant & 1\,, \label{eq:gm_red}\\
d^{*}_{1}[Q_{\mathrm{H}})](R_\mathrm{H}) & \leqslant & \sqrt{2\ln 2}\,, \label{eq:gm_reg_l1} \\
d^{*}_{\infty}[Q_{\mathrm{H}}](R_\mathrm{H})  & \leqslant & \tfrac{1}{2}\,,
\end{eqnarray}
where
\begin{equation}
R_\mathrm{H} = \log_2 |Q_{\mathrm{H}}| = m \log_2 m  + O\left(m\right)\,.  \label{eq:rate_h}
\end{equation}
\end{proposition}
\begin{proof}
We use Huffman code. Its KL-distance bound is well known~\cite{CoverThomas}. $L_{1}$ bound follows by Pinsker's inequality. $L_{\infty}$ bound follows from sibling property of Huffman trees~\cite{Gallager}.
It remains to estimate the number of Huffman trees $T_m$ with $m$ leaves.
Consider a skewed tree, with leaves at depths $1,2,\ldots,m-1,m-1$. The last two leaves can be labeled by $\binom{m}{2}$ combinations of letters, whereas the other leaves - by $(m-2)!$ possible combinations. Hence $T_m \geqslant \binom{m}{2}(m-2)! =\frac{1}{2} m!$.
Upper bound is obtained by arbitrary labeling all binary trees with $m$ leaves: $T_m < m! \, C_{m-1}$, where $C_{m-1}$ is the Catalan number. Combining both we obtain: $- \tfrac{1}{\ln 2} m  < \log_2 T_m - m\log_2 m <  \left(2-\tfrac{1}{\ln 2} \right) m$.
\end{proof}


\subsection{Comparison}
We present comparison of maximal $L_1$ distances achievable by tree-based and type-based quantization schemes in Figure~3. We consider cases of $m=5$ and $m=10$ dimensions. It can be observed that the proposed type-based scheme is more efficient and much more versatile, allowing a wide range of possible rate/distance tradeoffs.

\section{CONCLUSIONS}
The problem of quantization of discrete probability distributions is studied. It is shown, that in many cases, this problem can be reduced to the covering radius problem for the unit simplex. Precise characterization of this problem in high-rate regime is reported. A simple algorithm for solving this problem is also presented, analyzed, and compared to other known solutions.

\section{ACKNOWLEDGMENTS}
The author would like to thank Prof. K.~Zeger (UCSD) for reviewing and providing very useful comments on initial draft version of this paper. The author also wishes to thank Prof.~B.~Girod and his students V.~Chandrasekhar, G.~Takacs, D.~Chen, S.~Tsai (Stanford University), and Dr. R.Grzesz\-czuk (Nokia Research Center, Palo Alto) for introduction to the field of computer vision, and collaboration that prompted study of this quantization problem~\cite{CHoGv2}.


\end{document}